\documentclass{article}
\usepackage{color}
\usepackage{amsmath}
\usepackage{amsfonts}
\usepackage{graphicx}

\definecolor{color1}{rgb}{0.000,0.000,0.000}
\definecolor{color2}{rgb}{0.000,0.000,1.000}
\definecolor{color3}{rgb}{0.000,1.000,1.000}
\definecolor{color4}{rgb}{0.000,1.000,0.000}
\definecolor{color5}{rgb}{1.000,0.000,1.000}
\definecolor{color6}{rgb}{1.000,0.000,0.000}
\definecolor{color7}{rgb}{1.000,1.000,0.000}
\definecolor{color8}{rgb}{1.000,1.000,1.000}
\definecolor{color9}{rgb}{0.000,0.000,0.502}
\definecolor{color10}{rgb}{0.000,0.502,0.502}
\definecolor{color11}{rgb}{0.000,0.502,0.000}
\definecolor{color12}{rgb}{0.502,0.000,0.502}
\definecolor{color13}{rgb}{0.502,0.000,0.000}
\definecolor{color14}{rgb}{0.502,0.502,0.000}
\definecolor{color15}{rgb}{0.502,0.502,0.502}
\definecolor{color16}{rgb}{0.753,0.753,0.753}

\begin{document}

\title{An Explicit Isochronal, Isometric Embedding}
\author{Earnest Harrison\footnote{email: \tt earnest.r.harrison@comcast.net}}
\maketitle
\begin{abstract}

Certain semi-Riemannian metrics may be decomposed into a Riemannian part
and an isochronal part. We use this idea and an idea of Kasner
to construct a manifold in 6+1 Minkowski space with a well known metric.
The full embedding we display is isochronal 
which simplifies visualizing the properties of the manifold. 
\end{abstract}

It is well known that embeddings of manifolds are not unique, 
however an embedding can be a useful representation of the manifold 
that facilitates understanding its properties. For example, an image of a manifold
may be thought of as a coordinate-free way of representing a given metric.

We will be concerned 
with isometric embeddings defined by a set of equations in $n+1$ 
Minkowski space that define a particular 3+1 dimensional subspace with 
the desired metric tensor. These equations take the form \cite{Dirac}: 
\begin{equation}
\begin{aligned}
&g_{\alpha \beta } = y_{,\alpha }^{i} \,y_{,\beta }^{j} h_{ij} \\
&\alpha ,\,\beta \in \left\{ 0,\,\ldots 3\right\} \quad i,\,j\in \left\{ 0,\ldots
n\right\} \quad h_{ij} =diag\left( -1,\,1\ldots 1\right)
\end{aligned}
\end{equation}

An embedding is defined once the functions, $y^{i} $,
are identified, whose derivatives combine to produce the desired 
metric tensor. In the following, we wish to restrict 
our attention to embeddings that are isochronal. That is, embeddings 
where the (single) time-like dimension is a linear function of 
time.

\section*{An Isochronal Embedding Example}

The metric we shall consider is Schwarzschild's vacuum metric
expressed in isotropic coordinates\cite{Stephani}:
\begin{equation}
\begin{aligned}
ds^{2} &=-c^{2} \left( \frac{2r-m}{2r+m} \right) ^{2} dt^{2} +\left(
1+m/2r \right) ^{4} \left( dr^{2} +r^{2} d\Theta ^{2} \right) \\
d\Theta ^{2} &= d\theta^2 + cos^2\left(\theta\right) d\phi^2,
\label{eq:isotropic}
\end{aligned}
\end{equation}
which we embed in 6+1 dimensional Minkowski space, $\mathbb{Y}$, where $y^0$ is the isochronal dimension:
\begin{equation}	
\begin{aligned}
y^{0}& =c\,t  \quad \mathrm{(timelike)}\\
y^{1}& =2k\,m\,\sqrt{1-\left( \frac{2r-m}{2r+m} \right) ^{2} } \cos \left(
\frac{c\,t}{2k\,m} \right) \\
y^{2}& =2k\,m\,\sqrt{1-\left( \frac{2r-m}{2r+m} \right) ^{2} } \sin \left(
\frac{c\,t}{2k\,m} \right) \\
y^{3}& =\left( 1+m/2r \right) ^{2} r\cos \left( \phi \right) \cos \left(
\theta \right) \\
y^{4}& =\left( 1+m/2r \right) ^{2} r\sin \left( \phi \right) \cos \left(
\theta \right) \\
y^{5}& =\left( 1+m/2r \right) ^{2} r\sin \left( \theta \right) \\
y^{6}& =\int \sqrt{\left( 1+\frac{m}{2r} \right) ^{4} -\left( 1-\left(
\frac{m}{2r} \right) ^{2} \right) ^{2} -8k^{2} m^{3} \frac{\left(
m-2r\right) ^{2} }{r\,\left( m+2r\right) ^{4} } } \,dr
\end{aligned}
\label{eq:isotropic1}
\end{equation}
All of the functions remain real if $r\geq 0$ and $k^{2} \leq 27$.

\begin{figure}[htbp]
\includegraphics[width=1.0\textwidth]{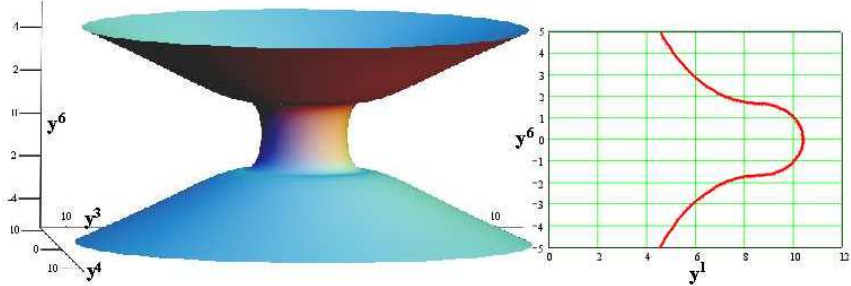}
\caption{Manifold slices $y^3 - y^4 - y^6$, and $y^1 - y^6$ at $t = \phi= \theta=0$.
The manifold rotates in the $y^1 - y^2$ plane.}
\label{fig:isotropic}
\end{figure}

Figure \ref{fig:isotropic} shows slices of this manifold for the critical case of $k^{2}=27$.
The $y^1 - y^6$ slice in this figure, appears as a line spinning in the $y^1 - y^2$ plane 
at a rate of $\omega =\frac{c}{2k\,m} $. The tip of the curve corresponds to
the event horizon, which is also at the waist of the $y^3 - y^4 - y^6$ slice.
This tip is spinning at the speed of light. That is,
the velocity vector becomes null at this point. The velocity of each point on the
curve models the time dilation we see in a gravitational field, which the
Minkowski space, $\mathbb{Y}$, takes care automatically.

The rotation of the manifold models acceleration. There are two components of
this acceleration: an in-manifold component and an out-of-manifold component.
The in-manifold component, which is in the $r$ direction, models the acceleration
due to gravity.\footnote{While it might seem that the acceleration modeled by a
rotation would be away from the mass, it is not. The acceleration is
toward the event horizon as can be seen in Figure \ref{fig:isotropic}.}

\section*{Isochronal Embedding Methodology}

Consider the line element for a static, spherically symmetric 
manifold:
\begin{equation}
ds^{2} =-c^{2}f^{2} \left( r\right) \,dt^{2} +g^{2} \left( r\right) \,dr^{2}
+h^{2} \left( r\right) \,d\Theta ^{2}
\label{eq:general}
\end{equation}
The first step is to define the isochronal, time-like function: 
\begin{equation*}
y^{0} =k_{1} \,c\,t,
\end{equation*}
 where $k_{1}$ is a free parameter. We then remove that term 
from (\ref{eq:general}) leaving a Riemannian sub-metric, which we then embed.

We use an idea due to Kasner \cite{Kasner} as carried forward by Fronsdal
\cite{Fronsdal}\footnote{The circular functions have the nice property of allowing us 
to include both $r$ and $t$ in these functions without creating 
unwanted off-diagonal terms in the metric tensor. Kasner used this property to represent
the Schwarzschild metric in six dimensions using two time-like dimensions. 
Fronsdal used hyperbolic functions and required only one 
time-like dimension. Neither of these embeddings were isochronal, however.}
to create the function, $f$, 
while compensating for the term introduced by $y^{0}$:
\begin{equation}
\begin{aligned}
y^{1}& =2k_{2} \,m\,\sqrt{k_{1}^{2} -f^{2} } \,\cos \left(
\frac{c\,t}{2k_{2} \,m} \right) \\
y^{2}& =2k_{2} \,m\,\sqrt{k_{1}^{2} -f^{2} } \,\sin \left(
\frac{c\,t}{2k_{2} \,m} \right)
\end{aligned}
\end{equation}
The term, $m$, is included in the circular function arguments 
to balance units.

We ensure that $y^{1}$ and $y^{2}$ remain real 
by demanding $k_{1}^{2} \geq f^{2} $.
The term, $k_{2}$, is a second free parameter.

The spherical term can be captured in the following three functions:
\begin{equation}
\begin{aligned}
y^{3}& =h\cos \left( \theta \right) \cos \left( \phi \right) \\
y^{4}& =h\cos \left( \theta \right) \sin \left( \phi \right) \\
y^{5}& =h\sin \left( \theta \right)
\end{aligned}
\end{equation}

So far, we have a metric tensor that defines a line segment as:
\begin{equation*}
ds^{2} =-c^{2}f^{2} \,dt^{2} +h^{2} \,d\Theta ^{2} +\left[ \left( h^{'} \right)
^{2} +\left( 2k_{2} \,m\,\sqrt{k_{1}^{2} -f^{2} } ^{'} \right) ^{2} \right]
\,dr^{2}
\end{equation*}
Where the prime denotes differentiation with respect to $r$. 
We will use $y^{6}$ to correct the coefficient of $dr^{2}$:
\begin{equation}
y^{6} =\int \sqrt{g^{2} -\left( h^{'} \right) ^{2} -\left( 2k_{2}
\,m\,\sqrt{k_{1}^{2} -f^{2} } ^{'} \right) ^{2} } \, dr
\end{equation}
The requirement that each of these functions remain real, places constraints 
on the two free parameters, $k_{1}$ and $k_{2}$.

Note that the manifold is dynamic, even though the metric tensor is static.
This is a general result: any isochronal manifold
that admits a metric that has nonvanishing gravitational field must be dynamic.
If it were not, there would be no acceleration due to gravity since the space-time 
we are embedding within is flat.

\begin{figure}[htbp]
\includegraphics[width=1.0\textwidth]{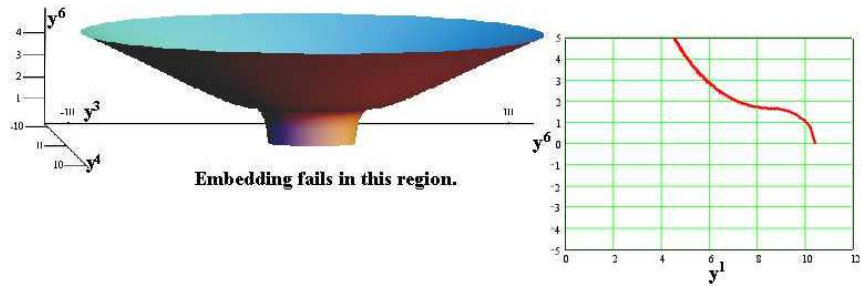}
\caption{The manifold defined using the Schwarzschild metric (\ref{eq:schwarzschild})
is identical to the one defined using (\ref{eq:isotropic1}) over the range of mutual validity.
This embedding method fails at the event horizon because of the coordinate singularity
in (\ref{eq:schwarzschild}).}
\label{fig:schwarzschild}
\end{figure}

As demonstrated above, the Isochronal Embedding Method is applicable 
to metrics that can be placed in diagonal form with bounded, 
real functions, $f$, $g$ and $h$, and with 
$g^{2} -\left( h^{'} \right) ^{2} -\left( 2k_{2} \,m\,\sqrt{k_{1}^{2}
-f^{2} } ^{'} \right) ^{2} $
everywhere non-negative for some fixed value of $k_{1}$ and $k_{2}$.
While these constraints exclude a global embedding of the Schwarzschild metric, 
we can find a piece of a manifold that has the Schwarzschild 
metric (the piece outside the event horizon):

\begin{equation}
\begin{aligned}
ds^{2}& =-c^{2} \left( 1-\frac{2m}{r} \right) \,dt^{2} +\frac{r}{r-2m}
dr^{2} +r^{2} d\Theta ^{2} \\
y^{0}& =c\,t \\
y^{1}& =2k\,m\,\sqrt{\frac{2m}{r}} \cos \left( \frac{c\,t}{2k\,m} \right) \\
y^{2}& =2k\,m\,\sqrt{\frac{2m}{r}} \sin \left( \frac{c\,t}{2k\,m} \right) \\
y^{3}& =r\cos \left( \phi \right) \cos \left( \theta \right) \\
y^{4}& =r\sin \left( \phi \right) \cos \left( \theta \right) \\
y^{5}& =r\sin \left( \theta \right) \\
y^{6}& =\int \sqrt{\frac{r}{r-2m} -1-2k^{2} \frac{m^{3} }{r^{3} } } \,dr 
\end{aligned}
\label{eq:schwarzschild}
\end{equation}
We can see that the two manifolds, (\ref{eq:isotropic1}) and (\ref{eq:schwarzschild}), are identical 
over their range of mutual validity -- just use the transformation to isotropic
coordinates\cite{Stephani}:
\begin{equation*}
r =\left(1+\frac{m}{2R}\right)^{2}R
\end{equation*}
We can also see this qualitatively in Figure \ref{fig:schwarzschild}.
We can see clearly that embedding (\ref{eq:schwarzschild}) fails at the
event horizon, while embedding (\ref{eq:isotropic1}), does not. This offers a way to analytically
continue a manifold in a natural way.\footnote{While isotropic coordinate transformation
``folds'' the functions $y^1$ - $y^5$ about the event horizon, it does not fold $y^6$, since $y^6$
is defined by an integral with a non-negative integrand.
Thus we have an analytic continuation, not a folding of the manifold.}

This example also shows that the manifold itself is coordinate-free. That is, the manifold image
(as seen in Figures \ref{fig:isotropic} and \ref{fig:schwarzschild}, in this case) is independent of
the coordinate system used to express the metric (over the range of validity of the embedding).

\section*{Summary}

We have demonstrated a method for decomposing certain semi-Riemannian metric
tensors into a Riemannian tensor and an isochronal tensor. We have shown that
we can embed the Riemannian portion in Euclidean space, $\mathbb{R}^n$, and then
add a single time-like isochronal dimension to that space to create a $n+1$ Minkowski space.
The resulting manifold, embedded in the Minkowski space, admits to the given metric tensor.
The resulting manifold provides a particularly simple way to visualize the force and
time dilation caused by gravity.

For this class of metric, the various embedding theorems that apply to Riemannian
manifolds can be extended to apply to semi-Riemannian manifolds. And since there
is considerable experience in  the mathematical community with Riemannian manifolds,
it is hoped that progress may be made applying other tools to metrics that 
admit to at least a local, isochronal embedding.


\begin{thebibliography}{9}

\bibitem{Dirac} Dirac, P.A.M., {\it General Theory of Relativity}, (John Wiley \& Sons, New York: 1975), p 11.
\bibitem{Stephani}  Stephani, H., {\it General Relativity, An Introduction to the Theory of the Gravitational Field}, (Cambridge University Press, New York: 1990), pp 113 - 114.
\bibitem{Dirac1} Dirac, P.A.M., op cit, pp 13, 14.
\bibitem{Kasner} Kasner, Edward, Am. J. Math. No. 43, p 130 (1921)
\bibitem{Fronsdal} Fronsdal, C., Phys Rev No. 116, p 778 (1959)

\end{thebibliography}
\end{document}